\documentclass{aastex62}
\usepackage{xcolor}
\usepackage[normalem]{ulem}
\usepackage{ulem} % for strikethrough

\reportnum{PASP 2018, v131, n995}

\usepackage[section]{placeins}

\newcommand{\BL}{{\textit{Breakthrough Listen }}}
\published{in PASP, 2018 Dec 4}
\shorttitle{Periodic Spectral Modulations}
\shortauthors{Isaacson et al.}

\begin{document}

\title{The Breakthrough Listen Search for Intelligent Life: No Evidence of Claimed Periodic Spectral Modulations in High Resolution Optical Spectra of Nearby Stars}

%\author[dddd-dddd-dddd-dddd]{Author Name}

\author{Howard Isaacson}
\affiliation{Astronomy Department, University of California, Berkeley, CA, USA}

\author{Andrew P. V. Siemion}
\affiliation{Astronomy Department, University of California, Berkeley, CA, USA}
\affiliation{ASTRON, Netherlands Institute for Radio Astronomy, Dwingeloo, NL}
\affiliation{SETI Institute, Mountain View, California, USA}

\author{Geoffrey W. Marcy}
\affiliation{Astronomy Department, University of California, Berkeley, CA, USA, Professor Emeritus}

\author{Jack Hickish}
\affiliation{Astronomy Department, University of California, Berkeley, CA, USA}

\author{Danny C. Price}
\affiliation{Astronomy Department, University of California, Berkeley, CA, USA}
\affiliation{Centre for Astrophysics \& Super-computing, Swinburne University of Technology, Hawthorn, VIC 3122, Australia}

\author{J. Emilio Enriquez}
\affiliation{Astronomy Department, University of California, Berkeley, CA, USA}
\affiliation{Department of Astrophysics/IMAPP, Radboud University, Nijmegen, The Netherlands}

\author{Nectaria Gizani}
\affiliation{Astronomy Department, University of California, Berkeley, CA, USA}

\email{hisaacson@berkeley.edu}

\begin{abstract}

We report on high-resolution spectra obtained by the Automated Planet Finder and high resolution optical Levy Spectrometer and the search for periodic spectral modulations, such as those reported in \cite{Borra2016}. In their analysis of 2.5 million spectra from the Sloan Digital Sky Survey, \cite{Borra2016} report periodic spectral modulations in 234 stars, and suggest that these signals may be evidence of extra-terrestrial civilizations.  To further evaluate this claim, we observed a total of three of the 234 stars with the Automated Planet Finder Telescope and Levy Spectrometer including all stars brighter than a visual magnitude of 14. Fourier analysis of the resultant spectra of these three sources does not reveal any periodic spectral modulations at the reported period, nor at any other period.
\end{abstract}



\section{Introduction}

\cite{Borra2016} reported spectral modulations with a period of $6.069 \times 10^{11}$~Hz in 234 stars with spectral types ranging from F2 to K1, observed with the Sloan Digital Sky Survey (Data Release 8, http://www.sdss3.org/dr8/).  They consider the possibility, as predicted in previous work \citep{Borra2010}, that the modulations in the spectra are caused by pulses of light generated by technological extraterrestrial sources. Using publicly available spectra provided by the Sloan Digital Sky Survey (SDSS) pipeline, \cite{Borra2016} used their own algorithms to search for modulations consistent with their hypothesized extraterrestrial signal. The SDSS spectra were obtained from two spectrographs, covering the blue (380--615 nm) and red (580--920 nm) regions.  They argue that the modulations are not instrumental, as only a small fraction of stars, $\sim 0.01 \%$, exhibit  modulation, and they are seen in both spectrographs, extending over the entire spectral range. 

 Their analysis uses a linear interpolation when converting the SDSS spectra from an assumed equally spaced wavelength sampling to equally-spaced frequency sampling. As only 234 out of 2.5M of the SDSS stars showed the modulation, they argue the effect is probably not instrumental. 

To independently test the existence of the reported periodic modulations, we obtained high resolution spectra of three of the stars reported by \cite{Borra2016}, and carried out a similar Fourier analysis of the spectra. These observations were undertaken as part of the Breakthrough Listen (BL) search for techno-signatures \citep{Worden2016}. BL is currently undertaking a targeted SETI survey \citep{Isaacson2017} that spans optical wavelengths with data from the Automated Planet Finder at Lick Observatory as well as radio frequencies with the 100m Green Bank Telescope (WV, USA) \citep{MacMahon2018} and the Parkes Telescope (New South Wales, Australia) \citep{Price2018}.  Since the \cite{Borra2016} claim analyzes optical spectra, our follow-up observations are exclusively from the Automated Planet Finder.

Previous optical SETI searches have searched for nano-second pulses at Harvard/Smithsonian Oak Ridge Observatory \citep{Howard2004}, pulsed signals  at Lick Observatory \citep{WrightS2001,Stone2005}, and laser lines in archived high resolution spectra, (\cite{Tellis2015}. These searches have placed limits on pulsed laser emissions on specific targets, including nearby stars, nevertheless, as the targets and methods differ, the results are not directly comparable to the results of \cite{Borra2016}, therefore we have collected and analyzed these observations. Michael Hippke has conducted an analysis of the SDSS spectra and searched for laser signals like those found by \cite{Borra2016}. While finding the same pulses in the SDSS spectra, he has not found them in spectra taken with a different telescope\footnote{https://github.com/hippke/laserpulses}.

In Section \ref{Observations} we discuss the Automated Planet Finder observations. In Section \ref{Fourier} we describe our search for modulated spectral signatures on the APF data.   Section \ref{conclusion} concludes the paper.

\section{Observations}\label{Observations}

On 2016 October 13, we used the APF-Levy to acquire high resolution optical spectra of three of the stars that are reported by \cite{Borra2016} to have periodic modulations in the spectra. The first star, TYC2037-1484-1,  has ICRS coordinates RA = (15\textsuperscript{h} 58\textsuperscript{m} 48.6\textsuperscript{s}) and DEC = (+27\degr 28\arcmin 03\arcsec), visual magnitude of 11.24, and spectral type G2. It is listed as observed on plate ID=3005 and fiber 270 in the SDSS.  The second star, TYC3010-1024-1, has ICRS coordinates RA = (11\textsuperscript{h} 04\textsuperscript{m} 19.8\textsuperscript{s}) and DEC = (+40\degr 10\arcmin 42\arcsec), Vmag=10.9, and spectral type F9. It is listed as observed on plate ID=3000 and fiber 71.  The APF exposure times are 20 minutes for both stars resulting in signal-to-noise ratio of 30:1 per pixel, which is 60:1 per resolution element. On 2018 April 18, we observed TYC2041-872-1, the only other star from the \cite{Borra2016} list that is brighter than V-magnitude = 14. The ICRS coordinates are RA = (16\textsuperscript{h} 01\textsuperscript{m} 33.3\textsuperscript{s}) and DEC = (+27\degr 33\arcmin 55\arcsec). It has a visual magnitude of 12.5, spectral type of F9 and is listed on plate ID 3005 and fiber 162. The exposure time is 15 minutes resulting in SNR equal to 22 per pixel. The noise in this type of optical spectra is dominated by Poisson errors, so the SNR is simply the square root of the average number of counts per pixel. We employed a decker size, or slit opening size that projects to  $1\farcs0 \times 3\farcs0$ on the sky. 

We examined all targets in Tables 1 and 2 from  \cite{Borra2016} that are observable  with the APF. A star must be brighter than V=14 in order for the guide camera to acquire the target \citep{Radovan2014}. We acquired spectra of the three stars brighter than V=14.0. SDSS spectra have typical exposure times of 45 minutes broken into three exposures\footnote{http://classic.sdss.org/dr7/instruments/spectrographs/index.html}, resulting in similar sensitivity for APF to detect the signal found in the SDSS spectra.

The APF-Levy spectra provide a high spectral resolution of $\lambda / \Delta \lambda$=95000  and span a wavelength interval of 374 - 950 nm \citep{Radovan2014}. In comparison, the SDSS spectra have an average resolution of $\lambda / \Delta \lambda$=2000 and span a wavelength interval of 380 - 920 nm. Thus, the APF-Levy spectra  have superior spectral resolution and comparable wavelength coverage to those used by \cite{Borra2016} to search for periodic modulations in the spectra. The modulations are periodic in frequency units (not wavelength units) with a period of 6.069x10$^{11}$ Hz.  The APF-Levy spectral resolution corresponds to an average frequency resolution of 5.74x10$^{9}$ Hz, two orders of magnitude finer than the reported period of the modulations, thus easily resolving them.  The 75 spectral orders of the APF-Levy spectrometer each typically span 8x10$^{12}$ Hz, long enough to encompass 11-13 full modulation periods.  Thus the APF-Levy spectrometer has higher resolution and adequate wavelength span within 79 spectral orders to detect the reported periodic modulation. Six orders are omitted due to SNR less than 10 per pixel.

The APF-Levy spectra are reduced in the standard way as with all \BL spectra \citep{Isaacson2017}, involving bias subtraction, flat-fielding, and extraction to 1-D spectra for each spectral order. To maintain a consistent wavelength solution, spectra are obtained with the echelle grating set to the same location within a pixel for all observations. The standard wavelength solution is determined from both thorium-argon spectra and iodine spectra.  We immediately convert the wavelengths to frequencies at each pixel, suitable for subsequent Fourier analysis as done in \cite{Borra2016}. Neither the systemic velocity of the star or the barycentric correction of the spectrum is corrected.

\begin{figure}
\epsscale{.8}
\plotone{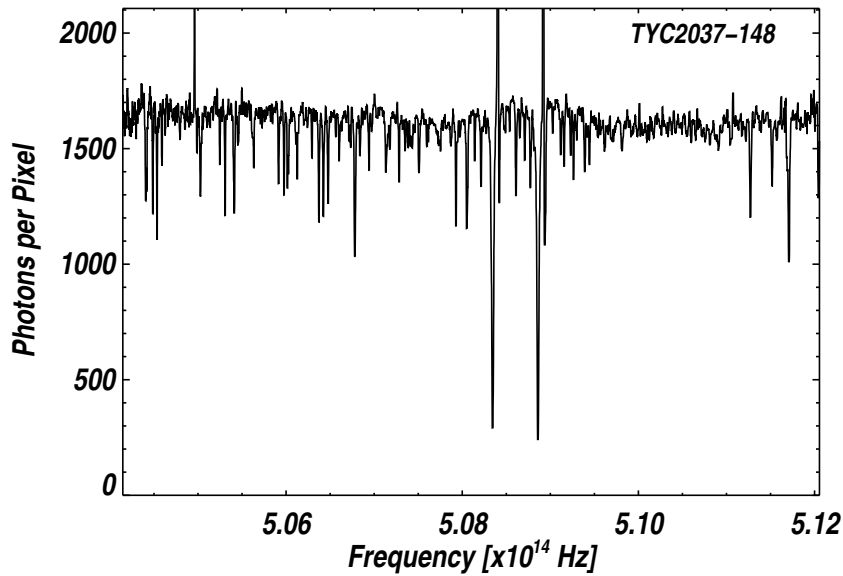}
\caption{Representative segment of the spectrum, in photons per pixel vs frequency, of TYC2037-148, that is reported to contain an artificial sinusoidal variation in intensity with frequency. The spectral resolution is R = 95000 and the signal to noise is 60 per resolution element.  This segment includes the Sodium D lines in absorption from the star. The two emission lines just longward are from the sodium street lights of San Jose and the spike at left is from a cosmic ray hitting the CCD.
}
\label{fig1}
\end{figure}

Similar to the normalization undertaken by \cite{Borra2016}, we remove the blaze function by dividing each spectral order by a median-smoothed (39 pixel) spectrum of a rapidly rotating B-type star, in this case HR7420.  The broad Balmer lines in absorption persist, and thus we incur false broad emission bumps a few percent above the spectral continuum near H-alpha, H-beta, H-gamma, H-delta, and H-epsilon. The resulting spectra have a flat continuum, preserving the resolution of 95000 and the signal to noise, typically 30:1 per pixel, or 60:1 per resolution element.  Figure \ref{fig1} shows a representative spectral order of one of the three target stars, TYC2037-148, near the Na D-lines.

In addition to testing blaze function removal with a spectrum of a rapidly rotating B-type star, we have experimented with several types of blaze function removal including using each individual spectra APF, as \cite{Borra2016} did with the SDSS spectra and using a quartz lamp spectrum. The choice of method for blaze function removal has no impact on our results.

\section{Fourier Analysis to Search for Modulations in the Spectra} 
\label{Fourier}
\subsection{Data Analysis}
We computed the Fourier power spectrum of the three spectra with the following procedure.  Operating separately on 70 spectral orders, encompassing wavelengths 385.3 - 920.8 nm, we used a cubic spline to rebin the spectrum to a frequency scale having a uniform frequency interval between adjacent pixels. The arbitrary frequency interval is 5.089847x10$^{14}$ Hz, corresponding roughly to two original pixels, thereby preserving spectral resolution within the Nyquist frequency. We computed the Fast Fourier Transform of the spectrum and took the Fourier modulus, as \cite{Borra2016} did.  This gave us 70 separate Fourier power spectra, one from each spectral order. 

Before searching for peaks, we combined the Fourier power spectra from each of the orders to form a final power spectrum. The spectral pieces within each spectral order have slightly different signal-to-noise ratios depending on the spectral energy distribution of the star and the efficiency of the Levy spectrometer. Rather than attempt to construct an optimal weighting algorithm we elected to simply sum the Fourier power spectra from all orders, giving equal weight to each.  While not optimal, we determine the sensitivity to period modulations by carrying out tests in which we inject synthetic modulations into the spectra, prior to computing the Fourier transform, to determine our sensitivity as a function of amplitude. If we were to see any hint (i.e. at 1 or 2-sigma) of the modulations, we would go back to optimize the summation of the Fourier analysis of the separate spectra orders.

The final summed Fourier power spectra from the three stars are shown in Figures \ref{fig2}, \ref{fig3}, and \ref{fig5}. The power spectra span the range of prospective periods in frequency from approximately $5 \times 10^9$ to $5 \times 10^{12}$~Hz, corresponding to the frequency interval of half the length of a spectral order to that of the spectral resolution.  The low values of power near 10$^{10}$ Hz shows that no coherent sinusoids are present in the spectra.  For longer prospective periods, the "noise" of absorption lines yields an increasing power which also shows no coherence, as evidenced by the lack of any distinct peaks. 

In Figures \ref{fig2}, \ref{fig3} and \ref{fig5} an arrow indicates the period of 6.07x10$^{11}$ Hz, the modulation reported in the SDSS spectra by \cite{Borra2016}.  We have found that none of the three stars are reported to have such a period show a peak with a significance above 1\% of the continuum intensity in the power spectra of the APF-Levy spectra here. {\it Thus, we cannot confirm the reported modulation periodicities for these three stars.}

\subsection{Calculation of Sensitivity}
We next determine the sensitivity of the APF-Levy spectra and our Fourier analysis to the presence of coherent periodicities in the spectra at the reported period of 6.09x10$^{11}$ Hz.  We took our APF-Levy spectrum of TYC2037-148 and multiplied the intensity vs frequency by a sinusoid of the form,

  $$W = 1 + A \sin{2 \pi ((\nu - \nu(0))/P_{BT})},$$

 where $A$ is the imposed amplitude of the sinusoid, $\nu$ is an array representing the frequencies at each pixel in the spectral order, $\nu(0)$ is the lowest frequency of the order, and $P_{BT}$ is the period of modulation, 6.0692244x10$^{11}$ Hz reported by \cite{Borra2016}.  Thus the spectrum within each spectral order is multiplied by a sinusoid centered on unity and having an amplitude, $A$ that is some fraction of the continuum intensity of the original spectrum. We ran trials with amplitudes, $A$, equal to 10\%, 3\%, 1\%, and 0.3\%. 
  
\begin{figure}
\epsscale{.8}
\plotone{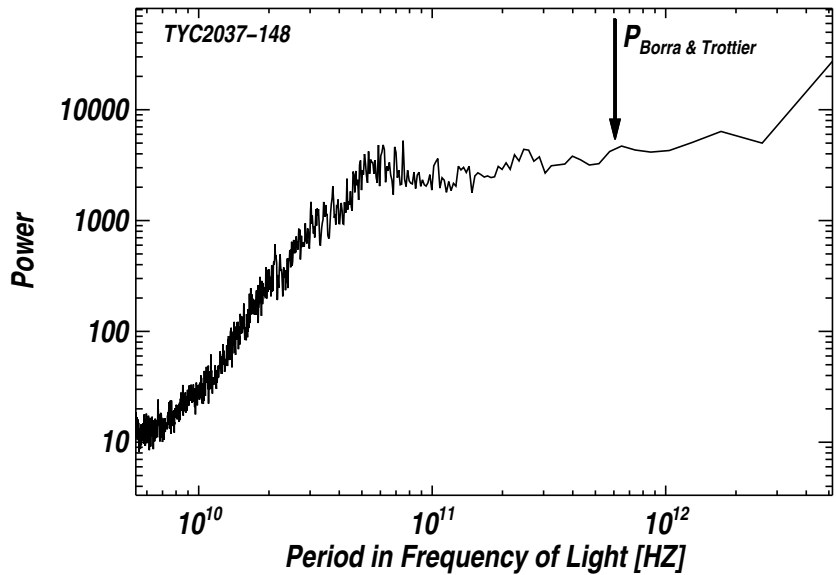}
\caption{The sum of the Fourier power spectra of 70 contiguous segments of the spectrum of TYC2037-148, spanning 3853 - 9208 Ang.
The horizontal axis represents increasing periods of the prospective periodicities in the intensity vs. frequency of the spectrum.
Any peak represents a periodicity in the intensity of light vs. frequency from the star.   No statistically significant peaks appear.
The period reported by \cite{Borra2016} 6.07 x 10$^{11}$ Hz is shown with the arrow.
}
\label{fig2}
\end{figure}

\begin{figure}
\epsscale{.8}

\plotone{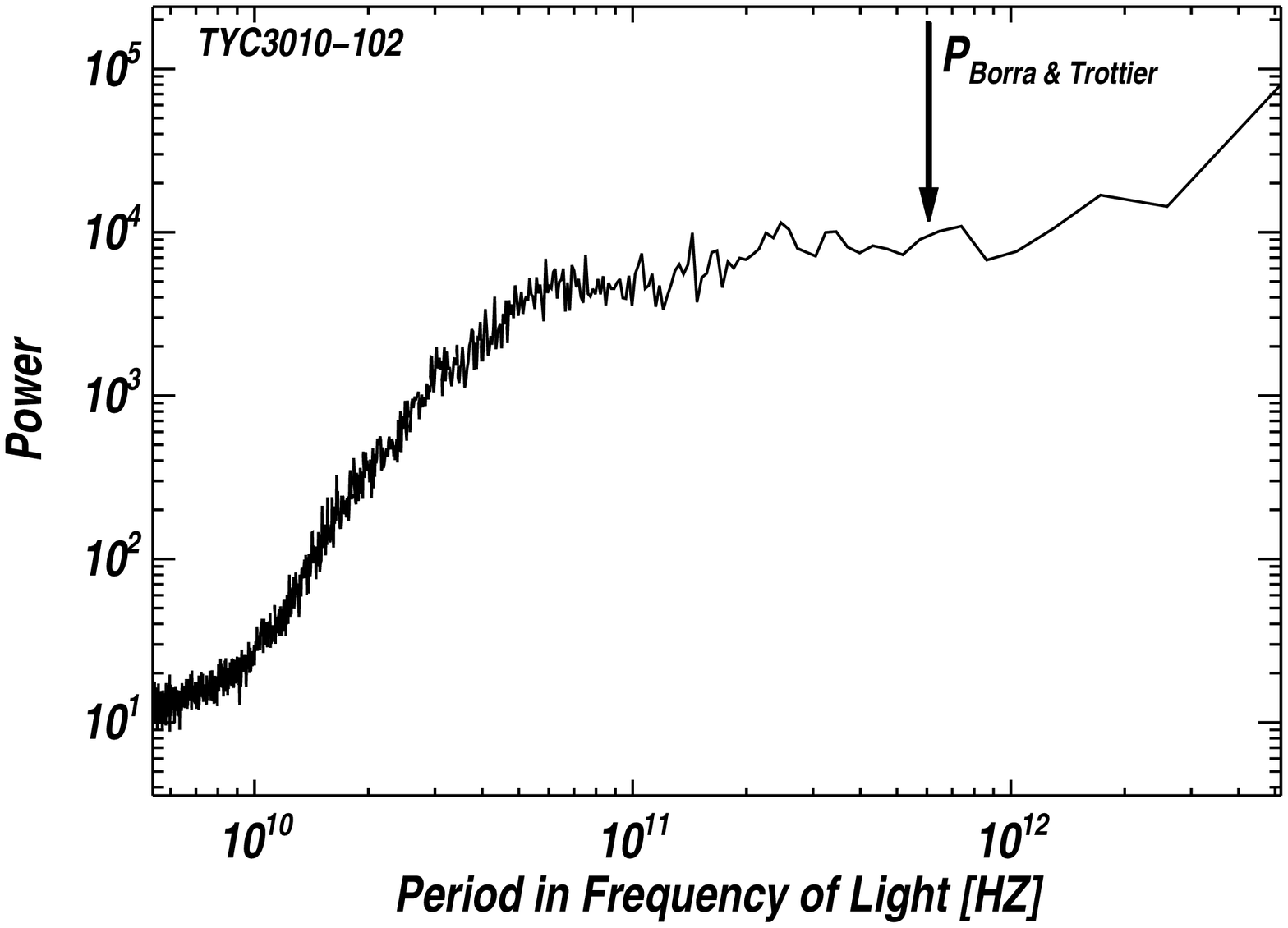}
\caption{Same as Figure \ref{fig2}, but this is the Fourier transform for TYC3010-102.  The period reported by \cite{Borra2016} is shown with the arrow.  No statistically significant peak appears in our spectrum.
}
\label{fig3}
\end{figure}

\begin{figure}
\epsscale{.8}
\plotone{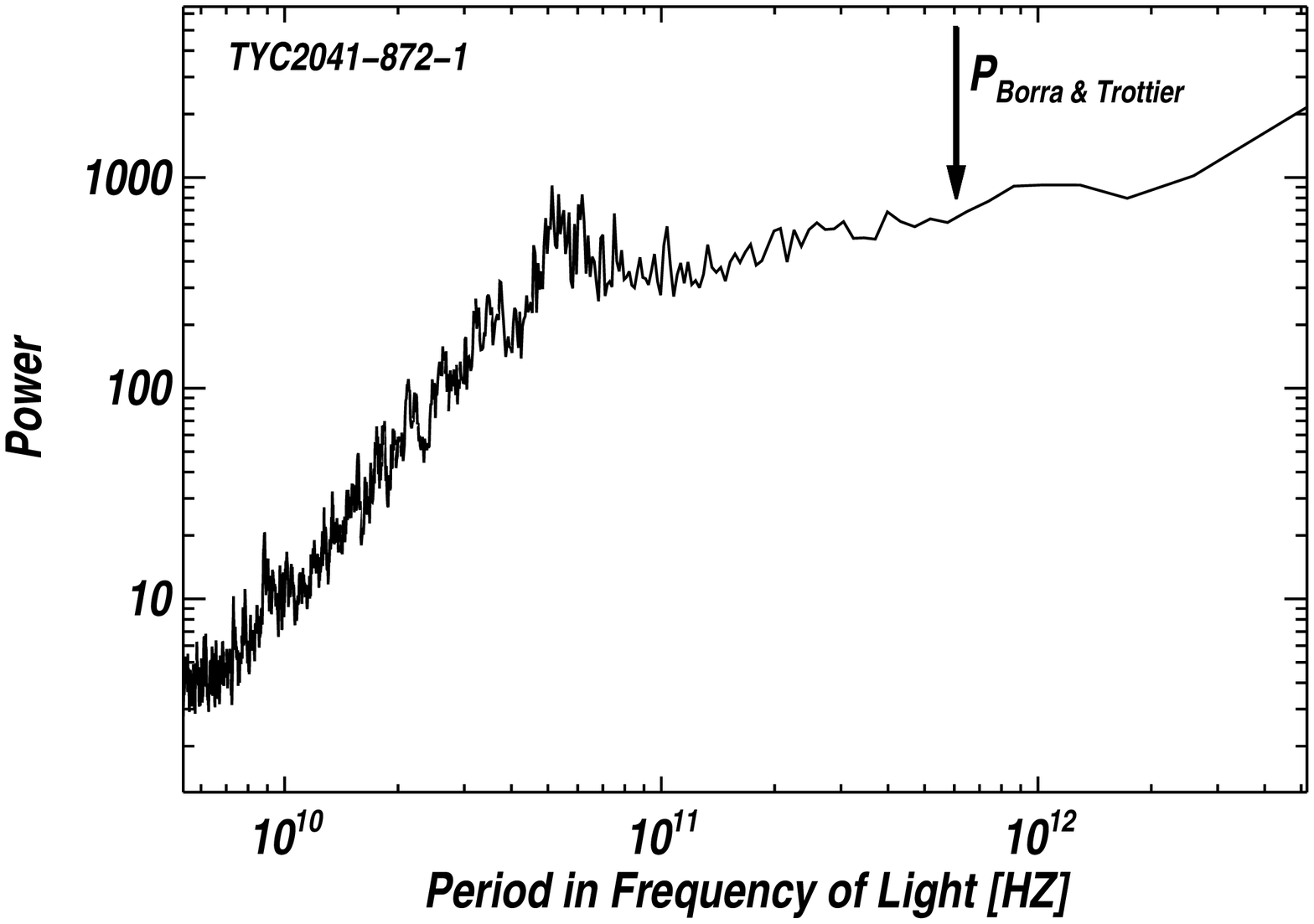}
\caption{Same as Figure \ref{fig2}, but this is the Fourier transform for TYC2041-872-1.  The period reported by \cite{Borra2016} is shown with the arrow.  No statistically significant peak appears in our spectrum.
}
\label{fig5}
\end{figure}

\begin{figure}
\epsscale{.8}
\plotone{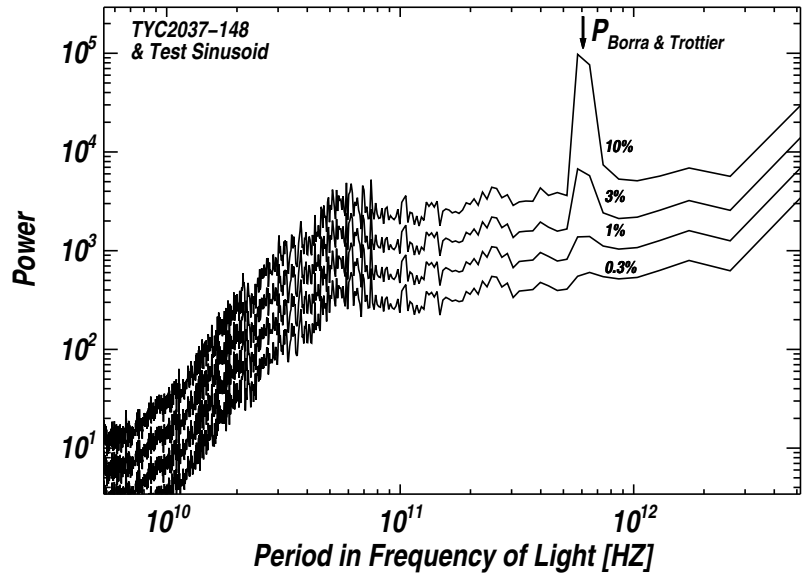}
\caption{The summed Fourier power spectra of TYC2037-148, but with an imposed synthetic periodicity in the spectrum at the period reported by \cite{Borra2016} of 6.069x10$^{11}$ Hz. Four different prospective amplitudes of the synthetic periodicity are imposed, 10\%, 3\%, 1\%, 0.3\%, of the continuum intensity of the spectrum. The power spectra are displaced vertically to allow all four to be seen.
The amplitudes of 10\% and 3\% appear clearly as a peak in the power spectrum, indicating their detectability. The amplitude of 1\% appears as a marginal peak, indicating marginal detectability. Amplitudes smaller than 1\% would not be detectable in our spectra.
}

\label{fig4}
\end{figure}

We operated on such spectra with the same algorithm as before, i.e. computing the summed Fourier power spectrum.  The result is shown in Figure \ref{fig4}. The spectra that had injected sinusoids with amplitudes of 10\% and 3\% of the continuum intensity clearly exhibit a peak at the injected period.  The spectrum that had an injected sinusoid with an amplitude of only 1\% of the continuum shows only a marginal peak in the power spectra, suggesting that such a 1\% amplitude would be detected only marginally.  Nonetheless, given the exact period value reported by \cite{Borra2016}  of such an exact period, the small peak height caused by the 1\% amplitude would certainly represent an intriguing coincidence, if seen in the actual spectra.  Such is not the case, as shown in Figures \ref{fig2}, \ref{fig3}, and \ref{fig5}. An amplitude of 0.3\% of the continuum intensity would not be detectable.  Thus we consider our detection threshold to be approximately 1\% of the continuum intensity. 
%\ee{This maybe discuss later, but it would be interesting to know to how much power this corresponds.}

The signal to noise threshold chosen by \cite{Borra2016} is 5.0. With the flux in our three APF spectra of 900, SNR = 30, and a sensitivity of 1\%, we are sensitive to signals with a flux of only 9, and SNR 3. Therefore our spectra have sufficient sensitivity to detect the claimed signal.

\section{Conclusions}\label{conclusion}

We obtained high resolution spectra of three of the stars that \cite{Borra2016} have reported as exhibiting a periodicity in intensity vs. frequency. Our spectra have comparable wavelength coverage and signal-to-noise ratio per SDSS resolution element, and have higher spectral resolution.  The Fourier spectra do not exhibit a coherent periodicity at any period, including that reported by \cite{Borra2016}. Thus, we do not find evidence of the periodic modulation in the APF spectra that is found by \cite{Borra2016} in the SDSS spectra for these three stars.

One possible explanation is that the laser-induced periodic modulation is not steady with time and was not operating when the APF spectra were taken. Another possibility is that the SDSS spectra contain some low level instrumental periodicity, at the few percent level, that happens only in some spectra.  Interferometric fringing can be sensitive to tiny changes, thermally or mechanically, in the optical configuration of the telescope and spectrometer. 

\acknowledgements{Acknowledgments: We thank Gloria and Ken Levy for support of the Automated Planet Finder Spectrometer. We gratefully thank Dan Werthimer for conversations about the spectral modulations. This work made use of the SIMBAD database (operated at CDS, Strasbourg, France) and NASA’s Astrophysics Data System Bibliographic Services. Funding for \BL research is sponsored by the Breakthrough Prize Foundation. Research at Lick Observatory is partially supported by a generous gift from Google.}

{\it {Facilities: Automated Planet Finder, Lick Observatory}}

\clearpage

\bibliographystyle{apj}
%\bibliography{Borra_Trottier}
% Replacing the bib file with the outputed .bbl file.
% Must update this if references added.
\bibliography{Borra_Trottier.bbl}

\begin{thebibliography}{}
\expandafter\ifx\csname natexlab\endcsname\relax\def\natexlab#1{#1}\fi

\bibitem[{{Borra}(2010)}]{Borra2010}
{Borra}, E.~F. 2010, \aap, 511, L6

\bibitem[{{Borra} \& {Trottier}(2016)}]{Borra2016}
{Borra}, E.~F., \& {Trottier}, E. 2016, \pasp, 128, 114201

\bibitem[{{Howard} {et~al.}(2004){Howard}, {Horowitz}, {Wilkinson}, {Coldwell},
  {Groth}, {Jarosik}, {Latham}, {Stefanik}, {Willman}, {Wolff}, \&
  {Zajac}}]{Howard2004}
{Howard}, A.~W., {Horowitz}, P., {Wilkinson}, D.~T., {et~al.} 2004, \apj, 613,
  1270

\bibitem[{{Isaacson} {et~al.}(2017){Isaacson}, {Siemion}, {Marcy}, {Lebofsky},
  {Price}, {MacMahon}, {Croft}, {DeBoer}, {Hickish}, {Werthimer}, {Sheikh},
  {Hellbourg}, \& {Enriquez}}]{Isaacson2017}
{Isaacson}, H., {Siemion}, A.~P.~V., {Marcy}, G.~W., {et~al.} 2017, \pasp, 129,
  054501

\bibitem[{{MacMahon} {et~al.}(2018){MacMahon}, {Price}, {Lebofsky}, {Siemion},
  {Croft}, {DeBoer}, {Enriquez}, {Gajjar}, {Hellbourg}, {Isaacson},
  {Werthimer}, {Abdurashidova}, {Bloss}, {Brandt}, {Creager}, {Ford}, {Lynch},
  {Maddalena}, {McCullough}, {Ray}, {Whitehead}, \& {Woody}}]{MacMahon2018}
{MacMahon}, D.~H.~E., {Price}, D.~C., {Lebofsky}, M., {et~al.} 2018, \pasp,
  130, 044502

\bibitem[{{Price} {et~al.}(2018){Price}, {MacMahon}, {Lebofsky}, {Croft},
  {DeBoer}, {Enriquez}, {Foster}, {Gajjar}, {Hellbourg}, {Isaacson}, {Siemion},
  {Werthimer}, {Green}, {Amy}, {Ball}, {Bock}, {Craig}, {Edwards}, {Jameson},
  {Mader}, {Preisig}, {Smith}, {Reynolds}, \& {Sarkissian}}]{Price2018}
{Price}, D.~C., {MacMahon}, D. H.~E., {Lebofsky}, M., {et~al.} 2018, ArXiv
  e-prints, arXiv:1804.04571

\bibitem[{{Radovan} {et~al.}(2014){Radovan}, {Lanclos}, {Holden}, {Kibrick},
  {Allen}, {Deich}, {Rivera}, {Burt}, {Fulton}, {Butler}, \&
  {Vogt}}]{Radovan2014}
{Radovan}, M.~V., {Lanclos}, K., {Holden}, B.~P., {et~al.} 2014, in \procspie,
  Vol. 9145, Ground-based and Airborne Telescopes V, 91452B

\bibitem[{{Stone} {et~al.}(2005){Stone}, {Wright}, {Drake}, {Mu{\~n}oz},
  {Treffers}, \& {Werthimer}}]{Stone2005}
{Stone}, R.~P.~S., {Wright}, S.~A., {Drake}, F., {et~al.} 2005, Astrobiology,
  5, 604

\bibitem[{{Tellis} \& {Marcy}(2015)}]{Tellis2015}
{Tellis}, N.~K., \& {Marcy}, G.~W. 2015, \pasp, 127, 540

\bibitem[{{Worden}(2016)}]{Worden2016}
{Worden}, P. 2016, in 67th International Astronautical Congress 2016, 34378

\bibitem[{{Wright} {et~al.}(2001){Wright}, {Drake}, {Stone}, {Treffers}, \&
  {Werthimer}}]{WrightS2001}
{Wright}, S.~A., {Drake}, F., {Stone}, R.~P., {Treffers}, D., \& {Werthimer},
  D. 2001, in \procspie, Vol. 4273, The Search for Extraterrestrial
  Intelligence (SETI) in the Optical Spectrum III, ed. S.~A. {Kingsley} \&
  R.~{Bhathal}, 173--177

\end{thebibliography}
\end{document}